# Large yield production of high mobility freely suspended graphene electronic devices on a PMGI based organic polymer

*Nikolaos Tombros,[1,2]\* Alina Veligura,[2] Juliane Junesch,[2] J. Jasper van den Berg,[2] Paul J. Zomer,[2] Magdalena Wojtaszek,[2] Ivan J. Vera Marun,[2] Harry T. Jonkman,[1] and Bart J. van Wees[2]*

[1] Molecular Electronics, Zernike Institute for Advanced Materials, University of Groningen, The Netherlands

[2] Physics of Nanodevices, Zernike Institute for Advanced Materials, University of Groningen, The Netherlands

\*e-mail: n.tombros@rug.nl

**The recent observation of fractional quantum Hall effect in high mobility suspended graphene devices[1,2] introduced a new direction in graphene physics, the field of electron-electron interaction dynamics. However, the technique used currently for the fabrication of such high mobility devices[3-5] has several drawbacks. The most important is that the contact materials available for electronic devices are limited to only a few metals (Au, Pd, Pt, Cr and Nb) since only those are not attacked by the reactive acid (BHF) etching fabrication step.[6] Here we show a new technique which leads to mechanically stable suspended high mobility graphene devices which is compatible with almost any type of contact material. The graphene devices prepared on a polydimethylglutarimide based organic resist show mobilities as high as 600.000 cm$^2$/Vs at an electron carrier density n = 5.0 10$^9$ cm$^{-2}$ at 77K. This technique paves the way towards complex suspended graphene based spintronic,[7] superconducting[8] and other types of devices.**

The widely accepted method to produce a suspended graphene electronic device is to deposit graphene on a silicon oxide (SiO$_2$) coated Si substrate,[9-10] contact the graphene layer with Cr/Au electrodes and to remove a part of the SiO$_2$ using wet etching with buffered hydrofluoric acid (BHF).[1-5] This technique has some serious drawbacks however. Since etching is isotropic, also the SiO$_2$ below the narrow Cr/Au contacts is partially removed[1-5] which can lead to mechanically unstable devices and a low yield of working devices. Also, the contact materials available for electronic devices are limited to Au, Pd, Pt, Cr and Nb since only those 5 metals are not attacked by BHF etching.[6] All other materials, metals (Al, Ti, Cu…), oxides (Al$_x$O$_y$, MgO, ITO…), ferromagnets (Co, Ni, Py…), half metals (CrO$_2$, LSMO...) and other anorganic materials necessary for the fabrication of graphene based spintronic,[7] superconducting,[8] optoelectronic and other types of devices are etched by BHF. Therefore, it is very crucial to find a method to produce a suspended graphene device which excludes this aggressive BHF etch step.

In this paper we present a unique technique which circumvents the above limitations and leads to mechanically very stable suspended graphene devices with a remarkably high yield. The idea is to deposit graphene on an organic polymer and to remove in a controllable way the polymer underneath the graphene layer using an organic solvent which is harmless to anorganic materials like metals, insulators and ferromagnets (Fig. 1). Multilayer resist technologies have previously been used for the fabrication of suspended carbon nanotube devices,[11,12] however those can not be used for the production of mechanically stable large size suspended graphene devices or with complex structures like suspended top gates and graphene patterned by etching. We used LOR-A (MicroChem), a polydimethylglutarimide (PMGI) based organic resist which shows excellent resistance to a wide range of solvents and is stable for temperatures up to 190°C. As we will show later, this technique not only is compatible with plasma etching techniques which can be used for example to etch a well defined Hall bar structure in graphene but also with techniques for the production of free standing top gates in order to obtain electrostatically defined free standing bilayer graphene quantum dots and graphene pn junctions. This technique can also be used for the production of a suspended graphene device on a flexible substrate for the investigation of mechanical strain related effects on the mobility and band gap of bilayer and single layer graphene.

The devices are prepared as shown in figure 1 (see **Methods**). Highly n-doped (0.007 Ωcm) 4" Si wafers covered with a 300 nm (or 500nm) silicon oxide dielectric are used as starting material on which we spin coat a 1.15 μm thick layer of LOR-A polymer. This thickness gives a good visibility of graphene in the green spectrum using an optical microscope. We deposit HOPG graphene on the LOR-A polymer using the Scotch tape technique[7] (Fig. 1a). Standard electron beam lithography (EBL) is used to fabricate a graphene electronic device (Fig. 1b,c). We evaporate 5 nm of Ti as adhesion layer and 75 nm of Au using an e-gun evaporator at 4.0 10$^{-7}$ mbar pressure (Fig. 1d). A very crucial step is lift-off (Fig. 1e) which is done in hot xylene (80$^o$C), an organic solvent which at this temperature is active enough to dissolve the PMMA layers but does not affect at all the (EBL exposed) LOR resist. To make graphene suspended, a second EBL step is used to expose the LOR resist underneath the graphene layer (Fig. 1f and 1g). We develop in ethyllactate to remove the EBL exposed LOR resist, rinse the sample in hexane and blow it dry gently with nitrogen.

All suspended graphene devices we made so far (26 in total) with lateral sizes up to 10 μm remain intact after this procedure and this remarkably without the use of a critical point drying system. There are three reasons for this. First the surface tension of hexane, $\gamma$ = 18nN/μm, is far too weak to rupture the suspended graphene layer during the drying process. The capillary force $F_c$ the solvent applies on a suspended graphene membrane having a width $W$, length $L$, suspended at a distance $d$ away from the substrate is[13] $F_c$ = 2W(L+d)/d $\gamma$ cosθ. Since hexane, the suspended graphene layer and any possible PMMA/LOR polymer contaminants on it are all hydrophobic, the wetting is very good, resulting to a contact angle of ~0$^o$. The capillary force applied on a $L$ = 10 μm, W = 1 μm suspended graphene layer is in this case around $F_{c0}$ = 0.4 μN, however the total (uniaxial) force needed to break a 1 μm wide graphene layer is approximately 42 μN,[14] which is more than two orders of magnitude larger. Second, since the distance between the suspended graphene and the SiO$_2$ substrate is 1.15 μm the chance that the graphene layer will stick on the SiO$_2$ surface by the van der Waals force is small as the graphene layer would first have to be stretched by $\varepsilon$ ~ 3% in order to reach the substrate. Taking into account that the elastic constant $E$ of graphene[14] is ~340N/m a strain of $\varepsilon$ = 3% asks for a capillary force of $F_c$ =

$2dWE\varepsilon/L$ = 2.2 μN >> $F_{c0}$. And third, each of the metallic electrodes connected to the graphene layer is supported by a solid pillar of LOR polymer (Fig. 1g) which makes the suspended graphene device mechanically very stable, not only against surface tension but also to the electrostatic force applied during gate voltage measurements. This last property is what lacks in large size suspended graphene devices produced by BHF etching of the $SiO_2$ substrate in which the part of the metallic electrodes on top of the graphene layer are not supported anymore by solid pillars of $SiO_2$ and can become therefore mechanically very unstable.[1-5]

Characterization of the electronic quality of the suspended graphene devices was performed using a standard lock-in technique by sending a current of 10 nA - 1 μA through the device in 2-probe and 4-probe geometries. All of our suspended devices show very low ohmic contact resistance (<60Ω) between the Ti/Au contacts and graphene, which is important for metrology purposes in order to obtain an accuracy of 1ppm or higher in the quantum resistance standard. Our suspended devices usually show strong p-doping and current annealing[15] is needed to remove the dopants originating from the PMMA and LOR polymer remains which cover the suspended graphene layer. The device shown in Fig. 2a-b (device **A**) was originally designed to test if it is possible to create a 10 μm suspended graphene layer without the use of a critical point drying system. Each part of the graphene layer between the Ti/Au electrodes was shown to be suspended after inspection under a scanning electron microscope. The graphene layer is suspended 1.15 μm above the 500 nm thick $SiO_2$ layer and has a gate capacitance of 10.5 aF/μm$^2$ (see below). The majority of the suspended regions shows strong p-doping except one (Fig. 2c) showing very low doping for which we extracted at T = 77K a mobility of 37.000 cm$^2$/Vs at a hole charge carrier density of $n_h$ = 4.7 10$^{10}$ cm$^{-2}$ (Fig. 2d). Current annealing (250 μA DC current) of the 4 μm long suspended part of the graphene device at T = 77K resulted in a 600.000 cm$^2$/Vs mobility device at an electron charge carrier density of $n_e$ = 5.0 10$^9$ cm$^{-2}$. Note that at this specific carrier density the first derivative of the resistivity curve versus gate voltage shows a maximum, which we interpret as the crossover into the metallic electron regime, where we can extract a well defined mobility.

The suspended graphene device (device **B**) shown in Figure 3a has a slightly lower capacitance of 8.6 aF/µm$^2$ , since it was prepared on a 300nm thick SiO$_2$ insulating layer coated with a 1.4 µm thick LOR-A resist. This capacitance is extracted from the 2-probe quantum Hall measurements in Fig.3d. Note that the geometrical capacitance for this device is 30% smaller than the one extracted from the quantum Hall filling factors. All mobilities presented in this work are therefore calculated using the electrical measured capacitance. Note that a similar discrepancy of the order of 15-30% between the geometrical and electrically defined capacitance were also found in other experimental studies.[1,5] On this device we performed a current annealing step in vacuum (10$^{-5}$ mbar) at room temperature (290K) by sending a 400µA DC current through the 3.2 µm wide suspended graphene which resulted in a slightly p-doped device with a mobility of 70.000 cm$^2$/Vs at n$_h$ = 2.2 10$^{10}$ cm$^{-2}$ (Fig. 3b and 3c). At T = 77K the mobility increased to 250.000 cm$^2$/Vs (n$_h$ = 1.0 10$^9$ cm$^{-2}$ ). In figure 3d we present the conductance of the device under a magnetic field applied perpendicular to the graphene layer at a temperature T = 4.2K. The appearance of the Landau plateau with the filling factor ν = 2 at a magnetic field of only 0.5T is a clear indication of the excellent electrical quality of the graphene layer. This together with the linear depedence of the filling factors 2, 6 and 10 on the gate voltage implies that possible sagging and/or deformation (e.q. due to electrostatic force from the gate electrode) is not important.[3]

We note that both devices presented in this paper, although exposed with 30keV electrons during the EBL exposure still show excellent conducting properties and high mobility, indicating that a 30keV low dose exposure does not introduce any significant damage to the graphene crystal. Note that we cannot exclude some (amorphous) carbon deposition due to electron beam exposure of the uncovered graphene. Since using current annealing we can obtain equally high mobilties as in processes which do not require electron beam exposure, we conclude that any deposited carbon, if already present, does not limit the electronic quality. We have also performed measurements on a suspended graphene device with 4 electrodes in a Hall geometry where we find a mobility of 380.000 cm$^2$/Vs (n$_e$ = 2.5 10$^9$ cm$^{-2}$, T = 4.2 K ) showing similar behavior as sample **B**. We have also succeeded in obtaining suspended graphene devices etched into a Hall bar geometry using a pure oxygen plasma in a reactive ion etching system (Fig. 4a) (manuscript in preparation). Such type of devices will be used for a more precise investigation of electron-electron dynamics in graphene. Since the PMGI based resist is sensitive to deep-UV light, standard optical lithography can be used for large scale preparation of suspended graphene devices (Fig. 4b). (manuscript in preparation)

In conclusion, we present a technique for the production of a suspended graphene device on a PMGI based resist which shows a superior behavior compared to the fabrication technique used currently by the graphene community in which a suspended graphene device is obtained after BHF etching of the $SiO_2$ substrate. First, the technique is compatible with all anorganic contact materials like metals, ferromagnets and insulators, which is not the case for the BHF etching technique in which only 5 contact materials Au, Pd, Pt, Cr and Nb are inert to the strong acid. Secondly, as long as the distance between suspended graphene and substrate is large enough, this technique does not require the use of a critical point drying system, an instrument which is a necessary tool for the production of 2 μm or longer suspended graphene devices when the BHF method is used.[1-5] And third, the mechanical stability is ensured in the LOR-A suspended devices since each of the electric contacts to the suspended graphene is supported by a solid column of polymer resist. The technology presented in this paper is also applicable for the production of suspended (carbon) nanotube/nanowire devices (with top gates), suspended submicrometer thin anorganic membranes and for the production of electrodes on a polydimethylglutarimide organic dielectric.

**Methods**

Highly n-doped (0.007 Ωcm) 4" Si wafers covered with a 300 nm (or 500 nm) silicon oxide dielectric are used as starting material. Standard optical lithography is used to produce gold markers on the SiO$_2$ surface which in a later stage will help to locate the dispersed graphene layers on the polymer layer. We spin coat a 1150 nm thick LOR-A layer and bake it on a hot plate at 200$^o$C for 15 minutes. This thickness gives a good visibility of graphene in the green spectrum using an optical microscope. We deposit graphene (HOPG grade YZA) on the polymer using the Scotch tape technique (Fig. 1a).[7] Electron beam lithography (EBL) is used to fabricate a graphene electronic device. In order to achieve a good undercut for succesfull lift-off of the EBL patterned stuctures we first spin coat a 300 nm thick 50K molecular weight polymethylmethacrylate (PMMA) layer (AllResist) which we bake on a hot plate at 180 $^o$C for 90s and afterwards we spin coat a 150 nm thick 450K PMMA layer (Elvacite 2041 in oxylene) and bake it on a hot plate for 90s at 180$^o$C (Fig. 1b). Note that the molecular weight of the PMMA and type of solvent is not important. The main requirement is that the top PMMA layer has a higher molecular weight with respect with the bottom PMMA layer and that both PMMA resists can be dissolved in hot xylene for the lift off process. The EBL exposure is done at 30 keV with an area dose of 180 μC/cm$^2$ and we develop the structures in xylene for 4 minutes at 21 $^o$C. We have chosen xylene since it does not develop the EBL exposed LOR resist. After development we rinse the sample in hexane and blow it dry with nitrogen (Fig. 1c). We evaporate 5 nm of Ti as adhesion layer and 75 nm of Au using an e-gun evaporator at 4.0 10$^{-7}$ mbar pressure (Fig. 1d). A very crucial step is lift-off (Fig. 1e) which is done in hot xylene (80$^o$C), an organic solvent which at this temperature is active enough to dissolve the PMMA layers but does not affect at all the (EBL exposed) LOR resist. Note that the LOR resist stays in the solid phase during lift-off since its glass transition temperature (190$^o$C) is well above 80 $^o$C. After lift-off we rinse the sample for 60s in ethyllactate at 21 $^o$C, 30s in hexane (21$^o$C) and blow dry with nitrogen. To make graphene suspended, a second EBL step is used to expose (area dose 1050 μC/cm$^2$) the LOR resist underneath the graphene layer (Fig. 1f and 1g). We develop in ethyllactate at 21$^o$C for 60s to remove the EBL exposed LOR resist, rinse the sample in hexane and blow it dry gently with nitrogen. Ultrasonic wire bonding is used to electrically contact the devices. To ensure good ohmic contact between the bonding wire and the contact material (Ti/Au) we introduced in addition a silver paste droplet (disolved in xylene) at the contact area.

To etch the Hall bar in Fig. 4a we introduced two extra fabrication steps in between steps e-f (Fig. 1). We spin coat ~100nm PMMA 400K disolved in oxylene on the sample, bake it on a hot plate at 180$^{o}$C, expose the parts of the graphene which have to be etched away with 30keV electrons (dose 180μC/cm$^2$) and develop for 4 min in oxylene. The Hall bar structure is etched using oxygen plasma in a reactive ion etching system.

**Acknowledgments**

We would like to thank Bernard Wolfs, Siemon Bakker and Johan G. Holstein for technical assistance. This work is part of the research program of the Foundation for Fundamental Research on Matter (FOM) and supported by NanoNed, NWO and the Zernike Institute for Advanced Materials.

**Author contributions**

N.T., A.V. and J.J fabricated the devices, performed the electronic measurements and analysed the data. J.J.B. contributed in the development of the technology. I.J.V.M. developed the current annealing method for our devices. M.W. and P. Z contributed in the etching step of the graphene Hall bar and optical characterization of the graphene layers. N.T, H.T.J. and B.J.W. supervised the experiments and analysis of the results and contributed to manuscript revision.

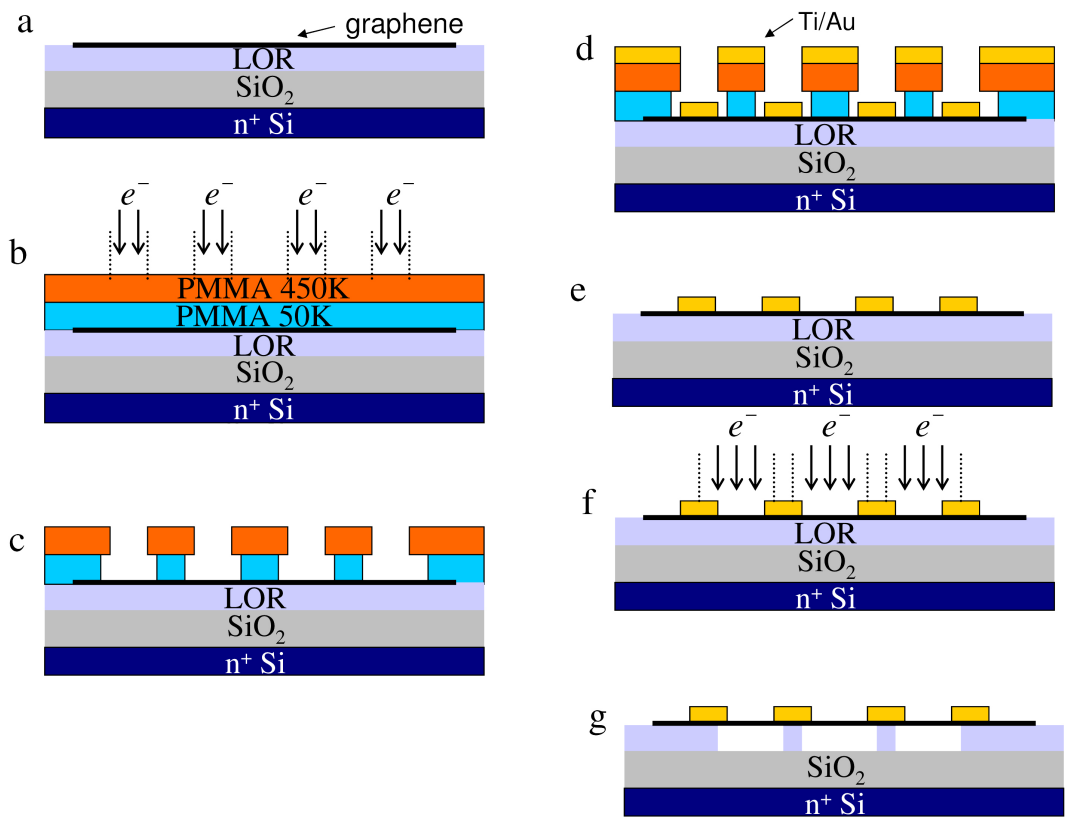

**Figure 1. Fabrication procedure of a suspended graphene device on LOR resist.** a) The scotch tape technique is used to deposit graphene on a LOR resist layer. b) For the electron beam lithography (EBL) step we spin coat two polymethylmethacrylate (PMMA) polymer resists on top of the graphene layer, the low molecular weight resist PMMA 50K and the high molecular weight resist PMMA 450K. Exposure is done at 30keV with an area dose of 180 μC/cm$^2$ c) Development of the exposed areas is done using xylene at 21°C. The undercut obtained in the EBL exposed structures is necessary for succesfull lift-off. d) Evaporation of Ti/Au. e) Lift-off is done in hot xylene (T=80 °C) f) The parts of the graphene layer which should be suspended are exposed with the EBL at 30keV and area dose 1050μC/cm$^2$. Note that instead of an EBL step also deep-UV exposure can be used to expose the LOR resist (see Fig. 4b). g) Suspended graphene is obtained after removal of the exposed LOR resist with ethyllactate developer at 21°C.

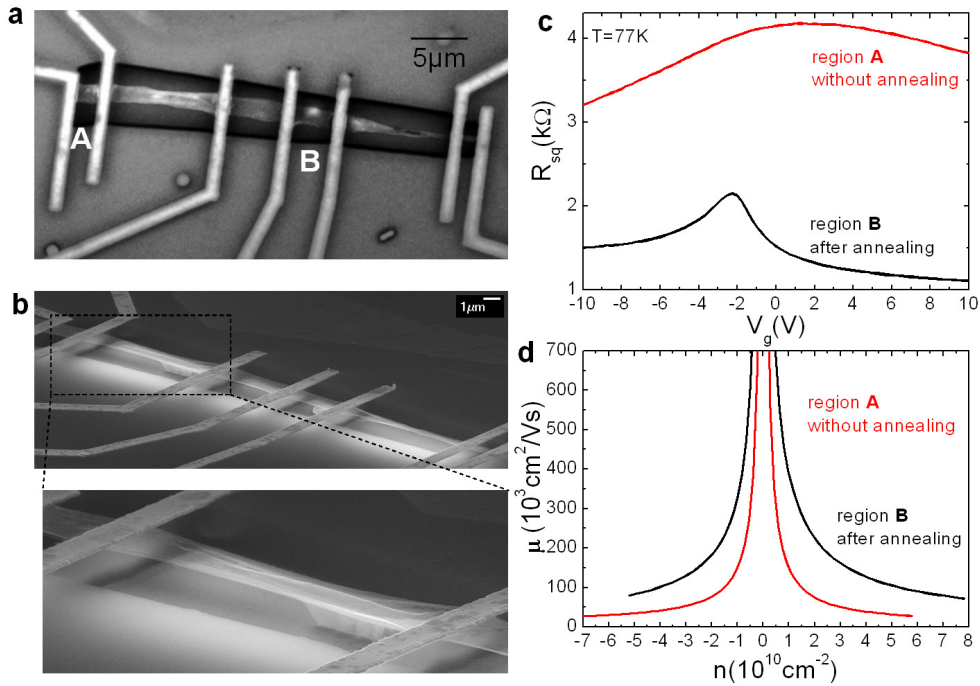

**Figure 2**. **Suspended graphene device on a LOR polymer layer.** a) Optical image of a 40 µm long graphene layer with 80nm thick Ti/Au electrodes and suspended graphene parts of (from left to right ) 2, 10, 5, 4, 10 and 2 µm. b) Picture taken with an scanning electron microscope under an angle of 70°. A zoom in at the 10 µm long suspended graphene layer shows a small amount of LOR-A resist underneath the graphene layer which was not removed away completely after 1 min of development in ethyllactate. This issue can be solved by increasing the development time from 1 min (used for this device) to 1.5 - 2 min. The disolution rate of the LOR layers is around 1.5 µm/min when exposed with 30keV electrons at an area dose of 1050µC/cm$^2$. c) The resistivity of the 2 µm long suspended graphene part (region A) showed already before an current annealing step a clear Dirac neutrality point at $V_g$ = 1V applied gate voltage. All other suspended graphene regions show strong p-doping and a current annealing step is needed to clean up the suspended graphene from polymer remains in order to observe the dirac neutrality point at ~0 V gate voltage. d) The mobility of the suspended graphene layer in region A is around 37.000 cm$^2$/Vs at $n_h$ = 4.7 10$^{10}$ cm$^{-2}$ for which no current annealing was used. Region B shows a mobility of 600.000 cm$^2$/Vs at $n_e$ = 5.0 10$^9$ cm$^{-2}$ after current annealing.

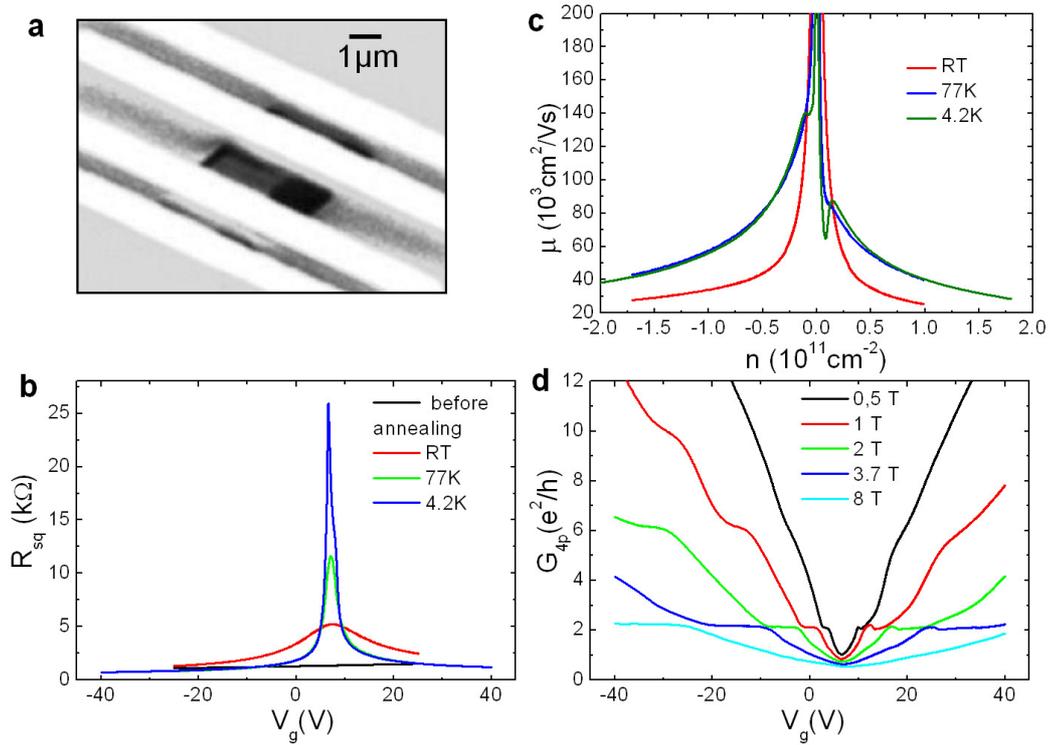

**Figure 3**. **Electronic measurements on a suspended graphene device prepared on a 1.4 μm thick LOR resist (300nm thick SiO$_2$).** a) Optical image of the sample. The outer Au electrodes are connected to a current source and the inner electrodes to a voltage probe. b) The device was initially heavily p-doped and after current annealing at room temperature (RT) in vacuum (10$^{-5}$mbar) we obtained a clear dirac neutrality point at V$_g$ ~8 V. The Dirac line width becomes very narrow at 77 K. The resistivity increases by a factor of 5 from RT to 4.2K. c) The RT mobility at n$_h$ = 2.2 10$^{10}$cm$^{-2}$ is 70.000 cm$^2$/Vs and increases to 250.000 cm$^2$/Vs at 77K and n$_h$ = 1.0 10$^9$cm$^{-2}$. d) Application of an external magnetic field B perpendicular to the suspended graphene layer at 4.2K. The characteristic quantum resistance plateaus for graphene at 2G$_0$, 6G$_0$ and 10G$_0$ are clearly visible at a magnetic field of 1T. The 2G$_0$ plateau is already well developed at B = 0.5T which is a clear indication of the exellent electronic quality of our suspended graphene device.

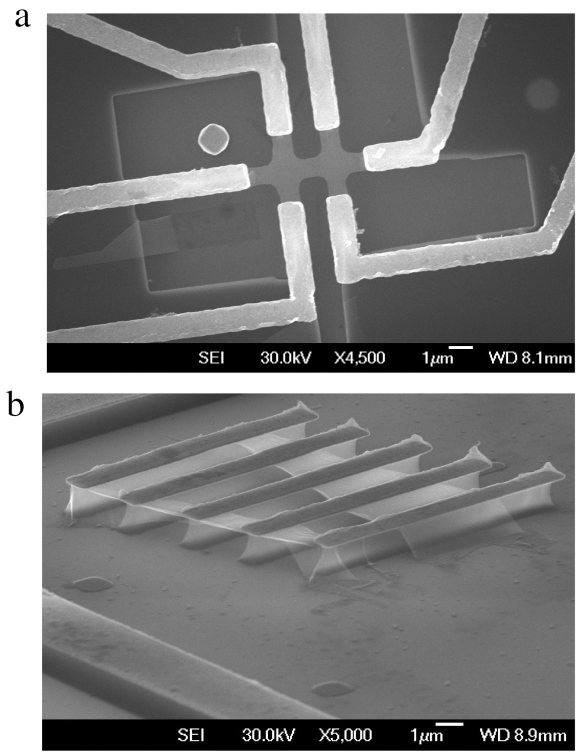

**Figure 4**. **Scanning electron microscope pictures showing the potential of the new technique** a) Top view of a freely suspended graphene Hall bar at 1.15 µm distance from the $SiO_2$ substrate. We introduced two extra fabrication steps in between steps e-f (Fig. 1) in order to etch a Hall bar (see **Methods**). b) Side view under 70º angle. Here, instead of the final EBL exposure, we introduced a short deep-UV exposure followed by development in ethyllactate to obtain suspended graphene (step f, Fig. 1). Note that the LOR resist below the gold contacts is not exposed since the 75nm thick gold functions as a mask layer for the deep-UV light. This shows the potential of using optical lithography for the mass production of suspended graphene devices farbicated with this technique.